\documentclass[12pt,preprint]{aastex}

\shorttitle{On the X-ray Baldwin effect}

\shortauthors{Jiang et al.}

\begin{document}

\title{On the X-ray Baldwin effect for narrow Fe K$\alpha$ 
emission line}

\author{P. Jiang, J. X. Wang, and T. G. Wang}
\affil{Center for Astrophysics, University of Science and Technology of China, 
Hefei, Anhui 230026, P. R. China \\
Joint Institute of Galaxies and Cosmology, USTC and SHAO, CAS}

\email{jpaty@mail.ustc.edu.cn}

\begin{abstract}
Most Active Galactic Nuclei (AGN) exhibit a narrow Fe K$\alpha$ line at $\sim$ 
6.4 keV in the X-ray spectra, due to the fluorescent emission from cold 
material far from the inner accretion disk. Using {\it XMM-Newton} observations, 
Page et al. found that the equivalent width (EW) of the narrow 
Fe K$\alpha$ line decreases with increasing luminosity
($EW \propto L^{-0.17\pm{0.08}}$), suggesting a decrease 
in the covering factor of the material 
emitting the line (presumably the torus). 
By combining the archival {\it Chandra} HETG observations of 34 type 1 AGNs with
{\it XMM} observations in literature, we build a much large sample with 101 AGNs.
We find a similar X-ray Baldwin effect in the sample 
($EW \propto L^{-0.2015\pm{0.0426}}$), however, we note that the anti-correlation 
is dominated by the radio loud AGN in the sample, whose X-ray spectra
might be contaminated by the relativistic jet.
Excluding the radio loud AGN, we find a much weaker anti-correlation 
($EW \propto L^{-0.1019\pm{0.0524}}$). We present Monte-Carlo simulations showing
that such a weak 
anti-correlation can be attributed to the relative short time scale variations
of the X-ray continuum.

\end{abstract}


\section{Introduction}
The iron K$\alpha$ emission line at $\sim$ 6.4 keV was first identified as 
a common feature in the X-ray spectrum of active galactic nuclei (AGN) by 
{\it Ginga} (Pounds et al. 1990; Nandra \& Pounds 1994). 
Its origin can be promptly interpreted as fluorescence emission following
photoelectric absorption of the hard X-ray continuum (see Reynolds \& Nowak 
2003 for a review). In 1995, {\it ASCA} observation of MCG--6-30-15 yielded a
broadened and skewed iron K$\alpha$ line profile (Tanaka et al. 1995).
The line profile was successfully interpreted as fluorescence line from the
innermost region of an accretion disk around a supermassive black hole.
This is the first direct evidence of emission from an accretion disk
extending down to a few Schwarzschild radii.
Thereafter, the iron K$\alpha$ emission line became one of the major targets
of X-ray astronomy studies.

Recent {\it XMM-Newton} and {\it Chandra} observations revealed narrow 
(unresolved by {\it XMM}) iron K$\alpha$ lines at $\sim$ 6.4keV in the X-ray 
spectra of most AGNs (e.g. Yaqoob \& Padmanabhan 2004; Page et al. 2004a).
Contrary to previous low-resolution {\it ASCA} observations,
broad iron emission lines appear to be much rarer in {\it XMM-Newton} and 
{\it Chandra} spectra. Different from the broad line profile, which comes
from the innermost region of the accretion disk, the narrow
iron K$\alpha$ is believed to be due to emission from neutral iron far from the inner
accretion disk. Possible origins of the narrow line include the outermost
regions of the accretion disk, the Broad Line Region, and the 
putative molecular torus.

In 1993, an X-ray Baldwin effect in the iron K lines was found in 
{\it Ginga} observations of 37 AGN (Iwasawa \& Taniguchi 1993). They found the 
equivalent width (EW) of the iron K line is anti-correlated to the
X-ray continuum luminosity (EW $\sim$ L$^{-0.20\pm0.03}_{2-10 keV}$).
The trend was confirmed by later {\it ASCA} observations of broad
iron lines (Nandra et al. 1997). Such an effect can be explained
due to the presence of an ionized skin on the accretion disk, 
with the degree of ionization increasing with luminosity (see
Nayakshin 2000a, b).

An X-ray Baldwin effect was also reported for the narrow 
iron K$\alpha$ line observed by {\it XMM-Newton} (Page et al. 2004a;
Zhou \& Wang 2005). Since the narrow line is produced much far away
from the inner X-ray continuum source, such an effect can not be due
to the change of the ionization state of the line emitting clouds.
A possible explanation is a decrease in the covering factor of the
material forming the fluorescence line (such as the torus) with
increasing X-ray luminosity. If true, such a result supports the
scheme that the lack of a large population of obscured quasars (luminous AGNs)
now discovered is due to the smaller covering factor of the torus in 
quasars (for example, the receding torus model, see Simpson 2005). 
However, the X-ray Baldwin effect of the narrow iron K$\alpha$ line
was questioned by {\it XMM-Newton} observations of PG quasars. 
Jim$\acute{\rm e}$nez-Bail$\acute{\rm o}$n et al. (2005) found the narrow iron K$\alpha$ lines in
PG quasars are similar to those of Seyfert galaxies, and no
X-ray Baldwin effect was found.
We also note that while some studies show that the fraction of 
obscured AGN significantly decreases with luminosity (e.g., Ueda et al. 2003;
Steffen et al. 2003), some other studies yield contrary results
supporting the existence of a large population of obscured quasars
(e.g. Mart\'\i nez-Sansigre et al. 2005; Tozzi et al. 2005).
Note a most recent work (Markwardt et al. 2005) studied the population
of obscured AGN in Swift/BAT all sky survey and found a significant
reduction in the fraction of absorbed/obscured AGN at higher luminosities.

In this paper we revisit this issue by studying archival AGN spectra obtained by
{\it Chandra} High Energy Transmission Grating Spectrometer (HETGS), which has
better spectral resolution in the iron line band for more accurate measurement
of the narrow Fe K line. We also combined our {\it Chandra} sample with {\it XMM} 
observations in literature to build a much larger sample of 75 radio
quiet and 26 radio loud AGNs. We found that 
the previous reported X-ray Baldwin effect is mainly due to the 
radio loud sources in their sample, whose X-ray spectra might be contaminated
by the relativistic jet. We found a much weaker X-ray Baldwin effect
of the narrow iron K$\alpha$ line for radio quiet sample, however, 
we present simulations showing that such a weak anti-correlation
is indistinguishable from an observational bias due to the variation
of AGN X-ray continuum.
Throughout this paper, H{\scriptsize 0} is taken to be 70 km $s^{-1}$ Mpc$^{-1}$, $\Omega${\scriptsize m}=0.27 and $\Omega${\scriptsize $\lambda$}=0.73.
    
\section{Chandra observations and data analysis}

We searched for archival {\it Chandra} HETGS observations as of 2005 
September 20 in the {\it Chandra} archive database\footnote{See http://cda.harvard.edu/chaser}.
The search revealed 89 HETGS observations of 
44 type 1 Seyfert galaxies and QSOs. Type 2 sources were excluded because
for most of them it is difficult to measure the intrinsic X-ray luminosities
due to strong absorption.
HETGS consists of two grating assemblies, a High Energy Grating (HEG)
and a Medium Energy Grating (MEG).
Our study will focus on the HEG data, which have better spectral resolution
and larger effective area in the Fe K$\alpha$ band. We also excluded
8 observations: MKN421 (Obsid 457), PG 1404+226 (812), 
1H 070-495 (862 and 2304), MCG--5-23-16 (2121-1e and 2121-2e), 
1H 0414+009 (2969), MRK 705 (4914), whose HEG count rates are 
too low for spectral fitting, and
3 high redshift QSOs (Q 0836+7104, PKS 2149-306, PKS 1830-211) because 
the center of their Fe K$\alpha$ lines in the observed frame were out of our 
restricted band (mentioned below).
Finally, we obtained 74 observations of 34 Seyfert 1 galaxies 
and QSOs from {\it Chandra}.

We used the first order of the HEG grating data by combining the 
positive and negative arms. For the sources with multiple observations,
we used the observation with the longest exposure time. We used
XSPEC version 11.3.2l (Arnaud 1996) for spectral fitting. To utilizing the 
highest possible spectral resolution available, we binned the spectra to the
intrinsic resolution of the HEG (one bin is $0.012\AA$), 
and adopt the {\it C}-statistic (Cash 1976) for minimization. 
Our model is a power-law plus a Gaussian emission line
with possible intrinsic neutral absorption. The Galactic absorption was 
also accounted.
We restrict our fit to 2.5 -- 8. keV band to avoid possible complex
absorption and/or soft excess at lower energy.  During the fit, the background 
was subtracted.

For \object{PDS 456}, \object{ARK 564}, \object{3C 273}, \object{PKS 2155-304}, 
\object{NGC 3227}, \object{MKN 421}, an Fe K 
emission line was not detected.  We obtained the upper limits on the 
equivalent width (EW) for such sources by fixing the line width to 1000 
km $s^{-1}$ and the rest frame line energy at 6.4 keV. 
For \object{MKN 766}, \object{3C 379}, \object{4c 74.56}, \object{1 ES 1028+511}, \object{IC 4329a}, \object{NGC 526a}, 
\object{NGC 985} and \object{EOS 198-G24} the narrow emission line was not resolved, we thus gave constraints on the line energy and EW by 
fixing the line width to 1000 $km s^{-1}$. 
For \object{NGC 7314} and \object{H 1821+643} obvious broad Fe K line can be 
detected with FWHM $\geqslant$ 10000 km $s^{-1}$, for which we also fixed the line width at 1000 km $s^{-1}$ to measure the line flux of the narrow core. 
For \object{MCG--6-30-15} whose Fe K line profile is very complex, we used 
the luminosity and narrow line equivalent width obtained by Yaqoob \& 
Padmanabhan (2004) directly. 
And for other unresolved objects (\object{MKN 590}, \object{MRC 2251-178}, \object{H 1821+643} and
\object{H 1426+428}) XSPEC could not give the best line energy while 
fixing the line width, so we also fixed the line energy.
For other sources, all three parameters of the Gaussian line
were set free during the fit. (See Table \ref{tbl-1}).

\section{Statistical analysis}

The Astronomy Survival Analysis (ASURV; Feigelson \& Nelson, 1985)
Package can be used in the presence of
censored (upper limit) data. This allows us to study
the relationship between line EW and L{\tiny 2-10keV},
including all significant and non-significant detections
of the narrow lines. We performed the Spearman Rank (SR) statistic to give
the correlation  between EW and luminosity in log-log space,
and performed the Buckley-James method to find the slope of the best-fit
line.
Figure 1a plots the rest frame EW of the narrow Fe K$\alpha$ line against the 
rest frame 2. -- 10. keV luminosity for {\it Chandra} sources. 
It can be seen that the EW decreases with the 
increasing luminosity. The fit gives a power law index of $\alpha=
-0.1940\pm0.0332$ (spearman's coefficient Rs = $-0.651$), where 
$EW\propto{\it L}^\alpha$. The result is consistent with $\alpha=-0.18\pm0.04$ 
obtained by Page et al. (2004a).
For our {\it XMM-Newton} observations\footnote{all of them were obtained from: 
Page et al. (2004a), Zhou \& Wang (2005), Jim$\acute{\rm e}
$nez-Bail$\acute{\rm o}$n et al. (2005), Page et al. (2004b)} 
(See Table \ref{tbl-2}), the fit gives 
a similar index of $\alpha=-0.2097\pm0.0503$ (Rs= $-0.463$) for the 
whole sample. 
However, when only the radio-quiet objects of the sample are considered 
(Figure 2b), the inverse correlation between EW and the
luminosity becomes weaker ($\alpha=-0.1023\pm0.0558$, Rs= $-0.280$).

To better study the possible correlation, we combined our {\it Chandra} sample
with our {\it XMM-Newton} sample to build a
much larger sample with 75 radio quiet sources and 26 radio loud sources.
There are 22 objects (\object{NGC 4151}, \object{NGC 5506}, \object{NGC 3516},
\object{NGC 4593}, \object{IC 4329a}, \object{MRK 509}, \object{F9},
\object{MRC 2251-178}, \object{PDS 456}, \object{NGC 3227}, \object{MCG-6-30-15},
\object{3C 273}, \object{MKN 766}, \object{3C 120}, \object{akn 564}, 
\object{NGC 4051}, \object{NGC 7314}, \object{Mrk 279}, \object{NGC 3783}, \object{ESO 198-G24}, 
\object{NGC 7469} and \object{NGC 5548}) included in both our {\it Chandra} and 
{\it XMM-Newton} samples. 
Figure 3a plots their {\it Chandra} X-ray luminosities against 
{\it XMM} ones. Clearly, 
there is no systematic bias between the observed luminosities by different
instruments, however, 8 out of 22 sources show variation with amplitude above 1.5.
The relationship of line EW observed by {\it Chandra} and {\it XMM} is plotted in Figure 3b.
We can see that most of EW observed by {\it Chandra} are consistent with those observed by {\it XMM}. Considering that {\it Chandra} data has better spectral resolution in the 
Fe K$\alpha$ band, which is essential to measure the narrow line, we directly
adopt {\it Chandra} data for these 22 sources and drop {\it XMM} ones in the 
combined sample.
Figure 4 plots 
the correlation between the EW and luminosity for all objects in 
the large sample ($\alpha=-0.2015\pm0.0426$, Rs = $-0.469$) and for 
radio-quiet objects only ($\alpha=-0.1019\pm0.0524$, Rs = $-0.266$). The fitting
results are consistent with those derived from {\it Chandra} and {\it XMM} data
 respectively. The values for three different samples are listed together 
in Table \ref{tbl-3}.
Note for the Chandra only sample, the difference between the correlation
index from RQ+RL sources and RQ sources only is not obvious as for XMM and
combined samples. This is mainly because that there is very limited number
of luminous sources in the Chandra only sample: only one RQ sources 
(H 1821+643) shows L$_X$ above 10$^{45.5}$ erg s$^{-1}$, thus the correlation
index is extremely sensitive to its EW measurement: for H 1821+643 we fixed 
the line width at 1000 km s$^{-1}$ to get an EW of 40 eV; while the best fit 
value gives a line width of 10000 km s$^{-1}$ and an EW of 140 eV; adopting 
the later values we get a linear correlation index of $-0.1062\pm0.0573$.

We can clearly see that when the RL sources were excluded from the sample the 
correlation between the line line EW and X-ray luminosity became much weaker
(with a confidence level less than 2$\sigma$). 
To show the results more clearly, in Figure 5 we plot the mean EW for 5
luminosity bins for all AGNs and just the radio-quiet objects in the 
combined sample. Following Page et al. (2004a), the upper limits 
were taken to be half of the value, together with 
an equally sized error bar. 
Similar to Page et al., we find a clear anti-correlation for RQ+RL
sources (Figure 5a). However, for RQ only sources, we see
no anti-correlation in Figure 5b.

\section{Time variation model}

Suggestions for the origin of the neutral Fe K$\alpha$ line include the 
putative molecular torus and/or the BLR (broad-line region). 
Since produced at a much larger scale, the line should exhibit no 
significant fluctuations within a timescale of months to years, while the
behavior of the X-ray continuum is much more active within much shorter 
timescales (Krolik et al. 1993). 
As a result, the observed equivalent width of the narrow line would be 
surely smaller while the X-ray continuum in a higher state, and vise versa.
For instance, Figure 6 plots the observations of \object{NGC 3783} and 
\object{NGC 4151} with {\it Chandra} (the only two radio quiet sources with
more than two {\it Chandra} HETG observations) and the best-fit lines. Although
there are only 6 points for NGC 3783 and 3 points for NGC 4151, we performed a
linear regression to give the best-fit slopes of them ($\alpha=-1.7752
\pm0.8130$, Rs = $-0.657$ for NGC 3783 and 
$\alpha=-1.5244\pm0.0732$, Rs = $-1.000$ for NGC 4151).
As expected, its EW clearly decreases with increasing luminosity. 
Such an effect may naturally lead to the observed anti-correlation slope 
between the EW and luminosity for a sample of AGN. In this section we
run Monte-Carlo simulations to check this possibility.

An anti-correlation between the X-ray luminosity and long term variation 
amplitude in Seyfert 1 galaxies was described by Markovitz \& Edelson (2004). 
Fractional variability amplitudes 
($F_{var}$) were measured for each light curve to quantify the 
intrinsic variability amplitude.
$$ F_{var}=\sqrt{S^2 - \langle \sigma_{err}^2 \rangle \over \langle X
\rangle^2} $$
where $S^2$ is the total variance of the light curve, $\langle
\sigma_{err}^2 \rangle$ is the mean error squared and $ \langle X \rangle
$ is the mean count rate of $N$ total points. They gave the 
anti-correlation between Fractional variability amplitude and luminosity: 
$F_{var} \propto L^{-0.135}_x$ for long (1296 days) timescale data.

Here we adopt a toy model to simulate the X-ray continuum variation, while
line flux is assumed to be invariable. The observed X-ray luminosity is 
assumed to be normally distributed with the width of the Gaussian distribution 
calculated to match the observed excess variance at different luminosities 
(Markovitz \& Edelson 2004).
By normalizing the observed line EW for the combined RQ sample (with upper 
limits) to the best-fit line, we first construct a set of line EW which 
does not correlate with the X-ray luminosity.
Random continuum variations are then added to the 
luminosities, and the line EW are modified correspondingly since we assume
no change to the line flux.
We repeated this step to build 1000 artificial samples with different random 
seed for each time. 
We used ASURV to perform linear regression to the artificial samples.
The distribution of the best-fit power-law slopes of the artificial samples
was presented in Figure 7.
We can see that 8.4\% of the simulations produce anti-correlation 
slopes steeper than the observed value, and the 
mean value is $-0.0485\pm0.0536$, which is compatible with the observed value 
($\alpha=-0.1019\pm0.0524$) within the errors.

\section{CONCLUSION}
In this paper, we studied the narrow iron K$\alpha$ line in 
123 Seyfert 1 galaxies and QSOs with archive {\it Chandra} HETGS observations 
and {\it XMM} observations in literature. 
There is an anti-correlation between the EW of the narrow, neutral iron line 
observed and the X-ray continuum luminosity: as the 2-10 keV rest frame 
luminosity increases, the 
equivalent width of the line drops. One possible reason for this 
negative correlation is a decline in the covering factor of the 
putative molecular torus as the 
luminosity increases (Page et al. 2004a). However, because of the dilution 
effect of relativistic beaming in the radio loud sources, they 
should be excluded from the study. 
Although Page et al. also reported an X-ray Baldwin effect after 
excluding RL AGN from their sample, the effect appears to be much
weaker for RQ AGN in our larger sample. 
We find that the observed weak anti-correlation
can be explained by the observational bias cause by variable AGN X-ray 
continuum.

\acknowledgments
This work was supported by Chinese NSF through NSF10473009/NSF10533050 and 
the CAS "Bai Ren" project at University of Science and Technology of China.

\clearpage

\begin{figure}
\plottwo{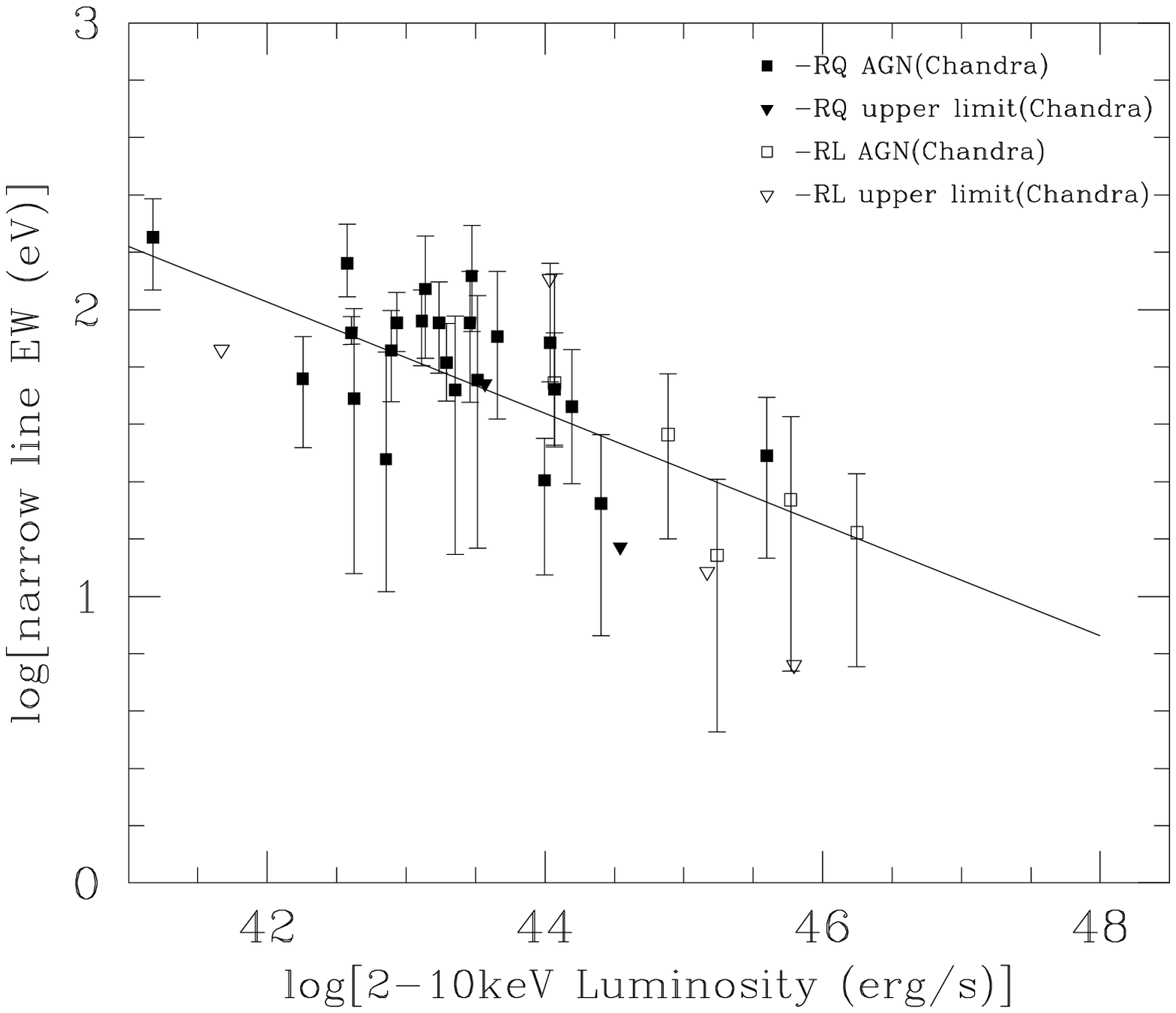}{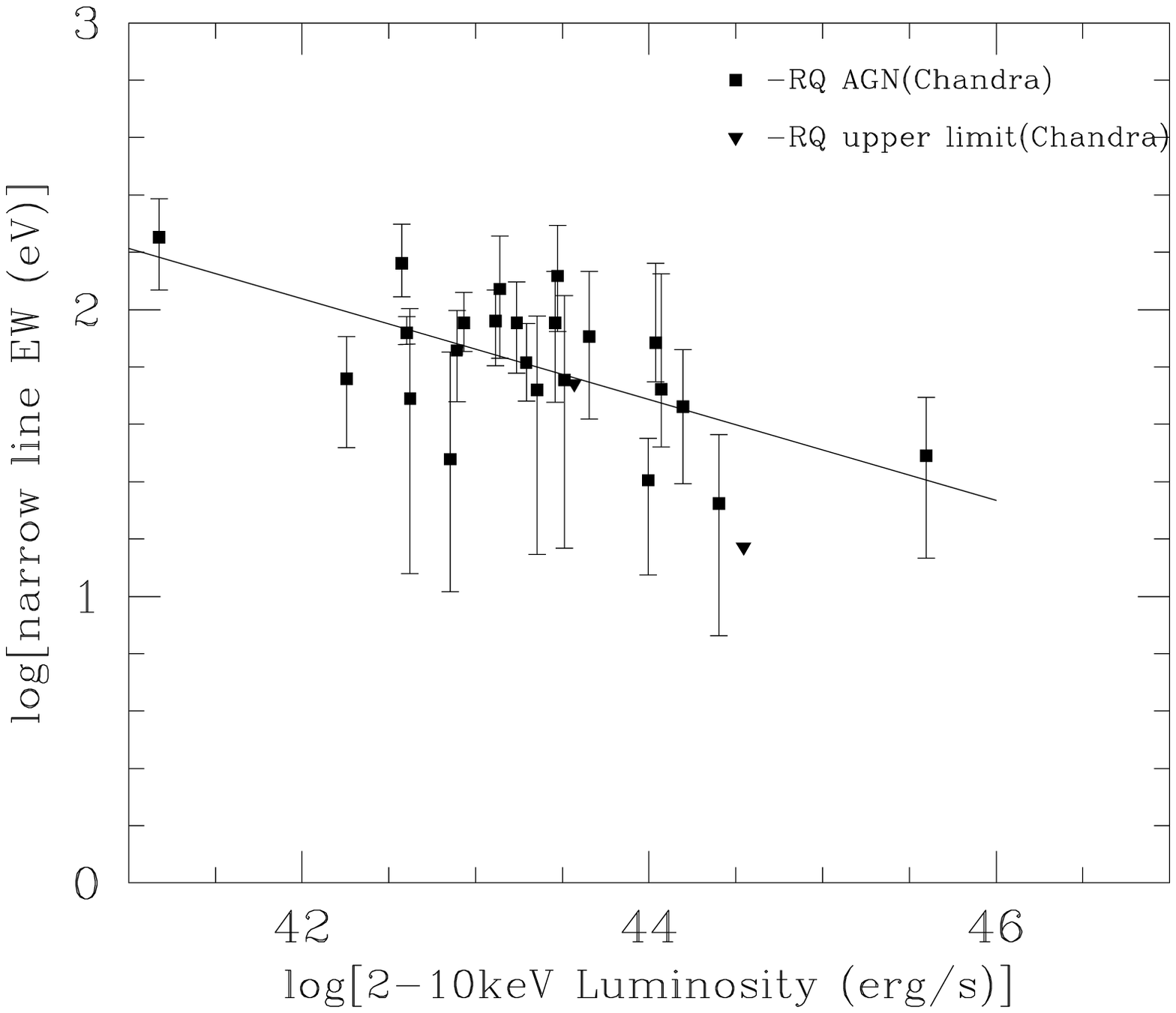}
\caption{The correlation between narrow FeK line EW and luminosity 
for the sample of 34 Seyfert 1 galaxies and QSOs observed by {\it Chandra}, 
Left: including both radio loud objects and radio quiet objects;
Right: including radio quiet objects only. The solid lines show the
best-fit anti-correlation slopes.
\label{fig1}}
\end{figure}

\clearpage

\begin{figure}
\plottwo{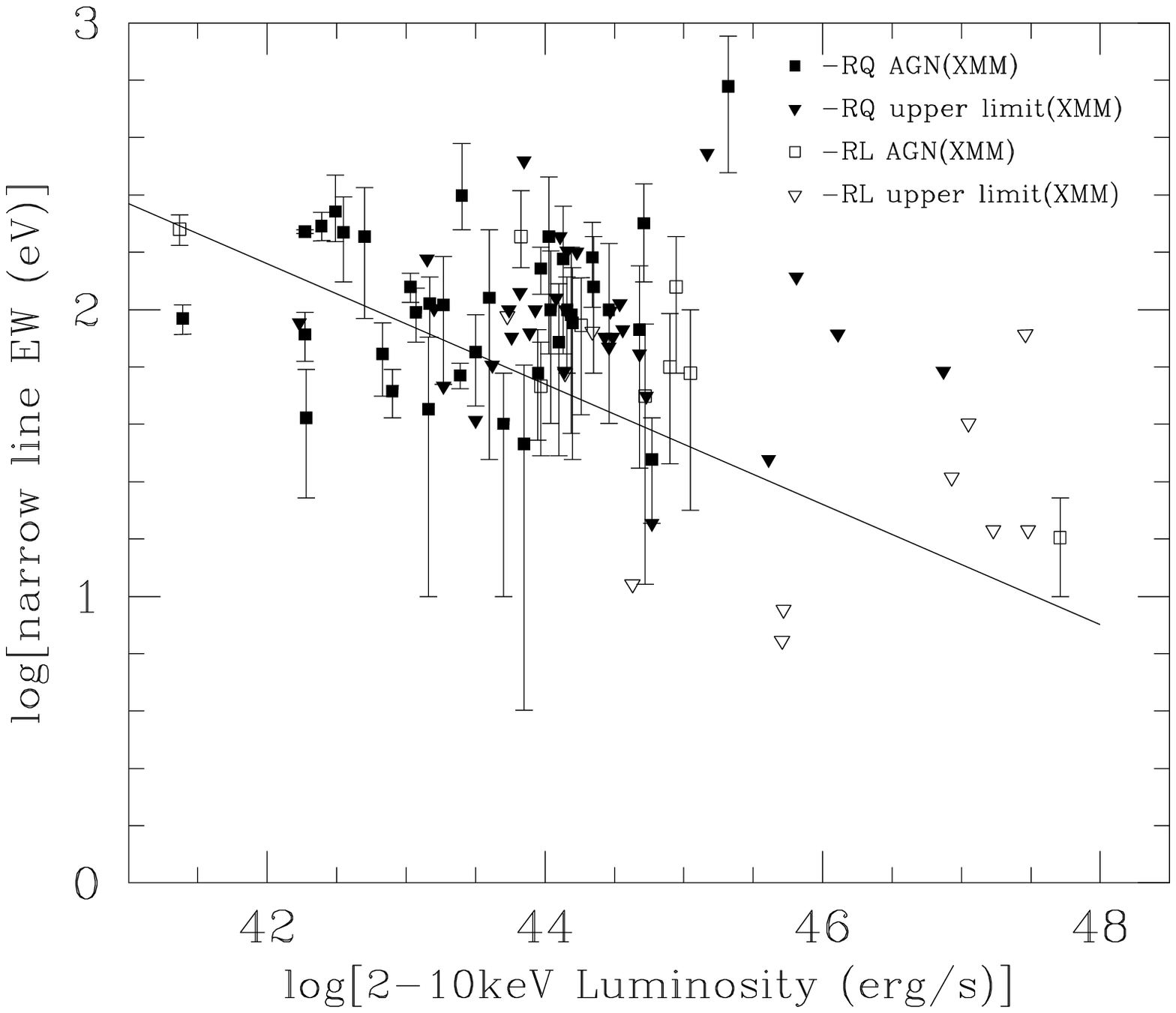}{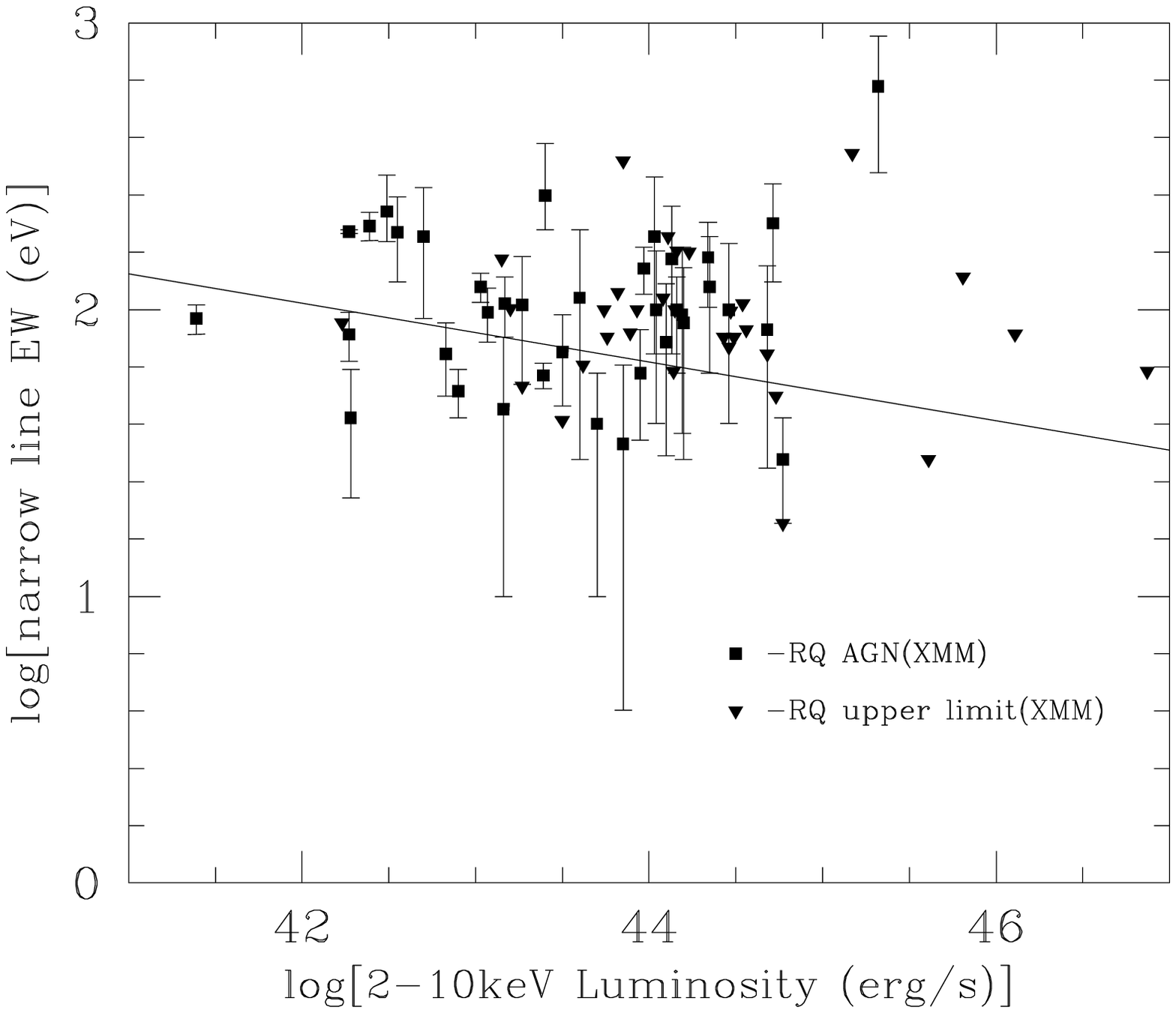}
\caption{The correlation between narrow FeK line EW and luminosity
for the {\it XMM-Newton} sample
Left: including both radio loud objects and radio quiet objects;
Right: including radio quiet objects only. The solid lines show the
best-fit anti-correlation slopes.
\label{fig2}}
\end{figure}

\clearpage

\begin{figure}
\plottwo{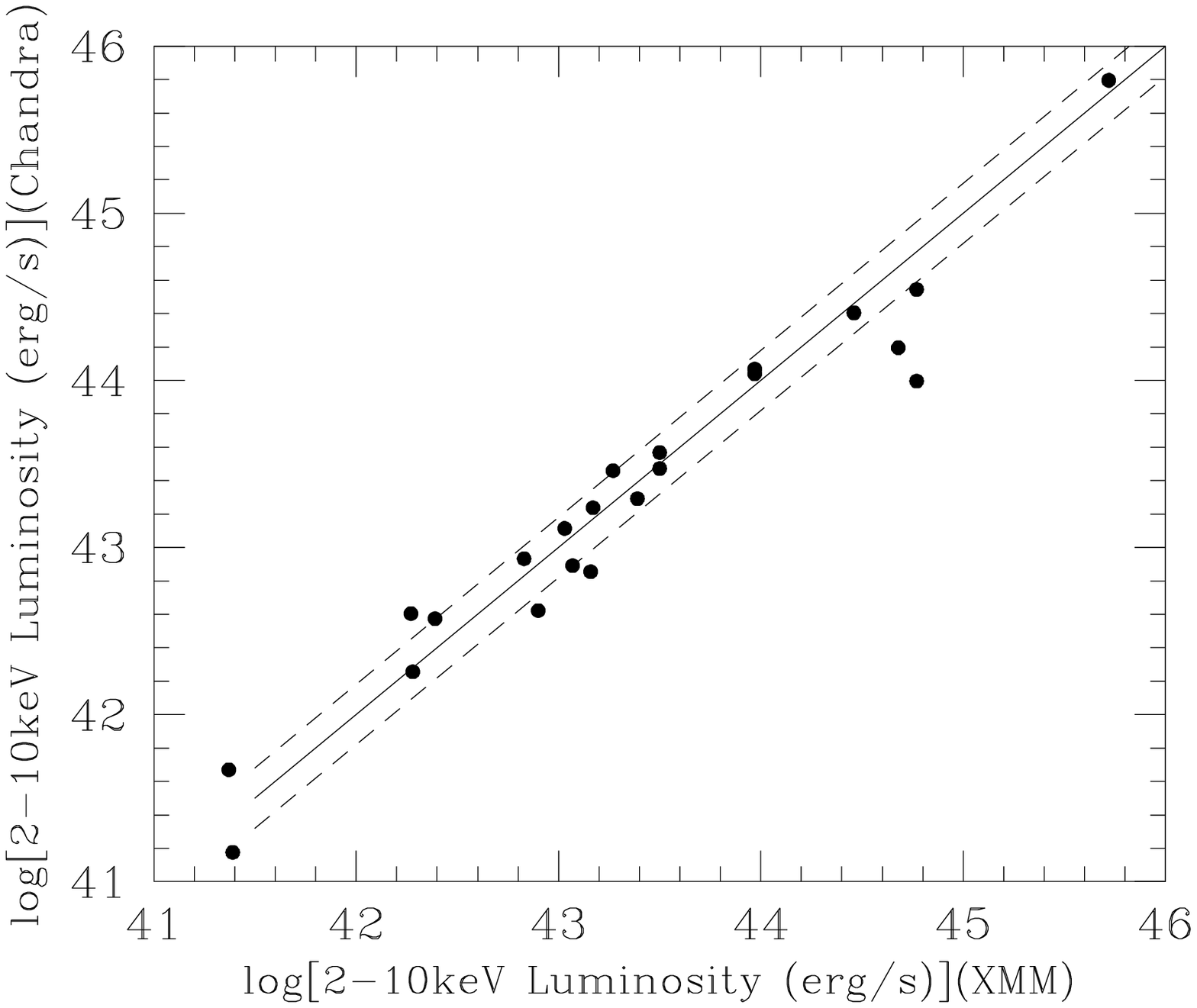}{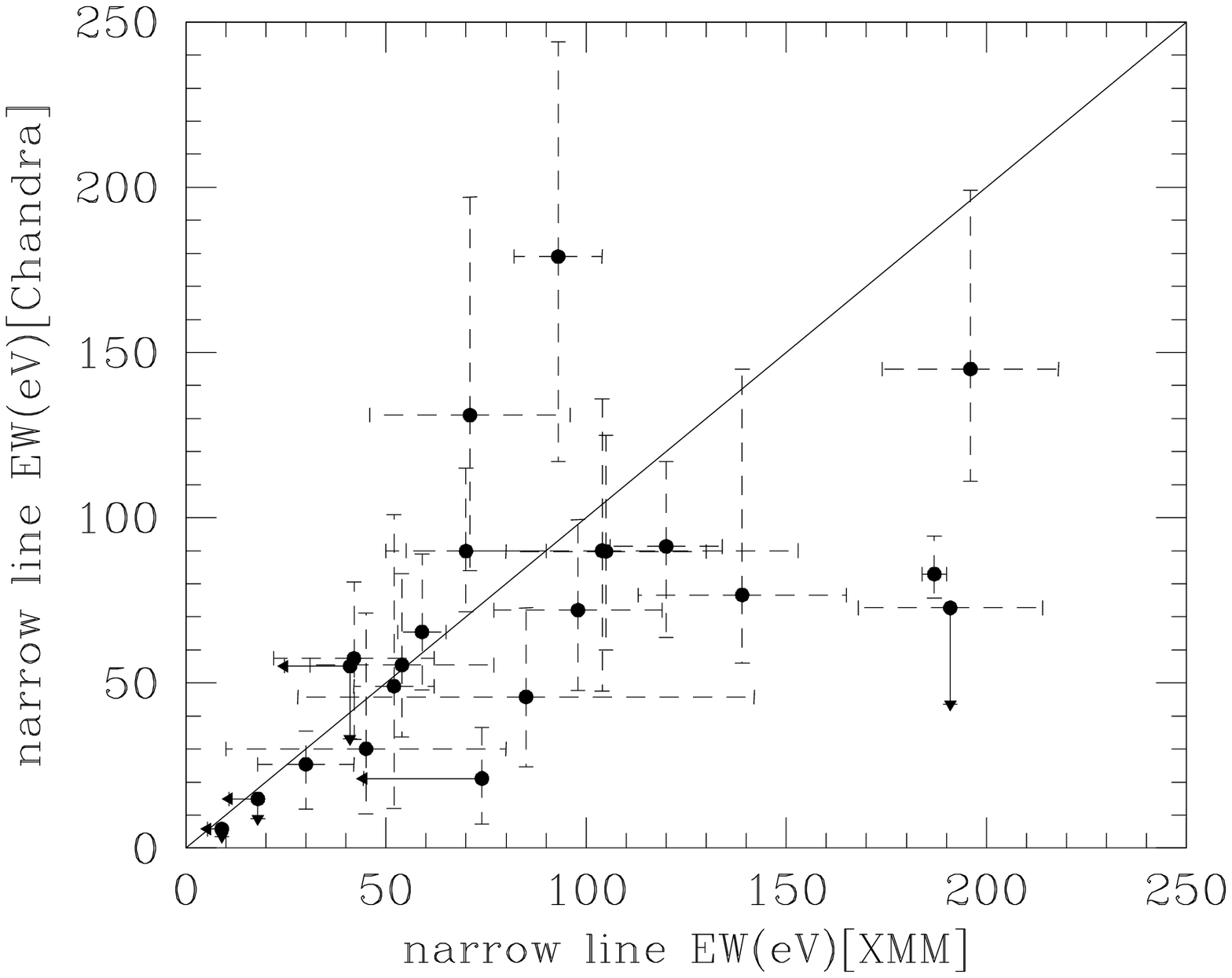}
\caption{Left: The luminosities obtained by {\it XMM-Newton} and {\it Chandra} 
for the 22 objects observed by both observatories. 
Solid line indicates $log[L_{XMM}]=log[L_{Chandra}]$;
dash lines indicate a variation amplitude of 1.5; Right: The EW 
observed by {\it XMM-Newton} and {\it Chandra} for the 22 objects.
Solid line indicates $EW_{XMM}=EW_{Chandra}$.
\label{fig3}}
\end{figure}

\clearpage

\begin{figure}
\plottwo{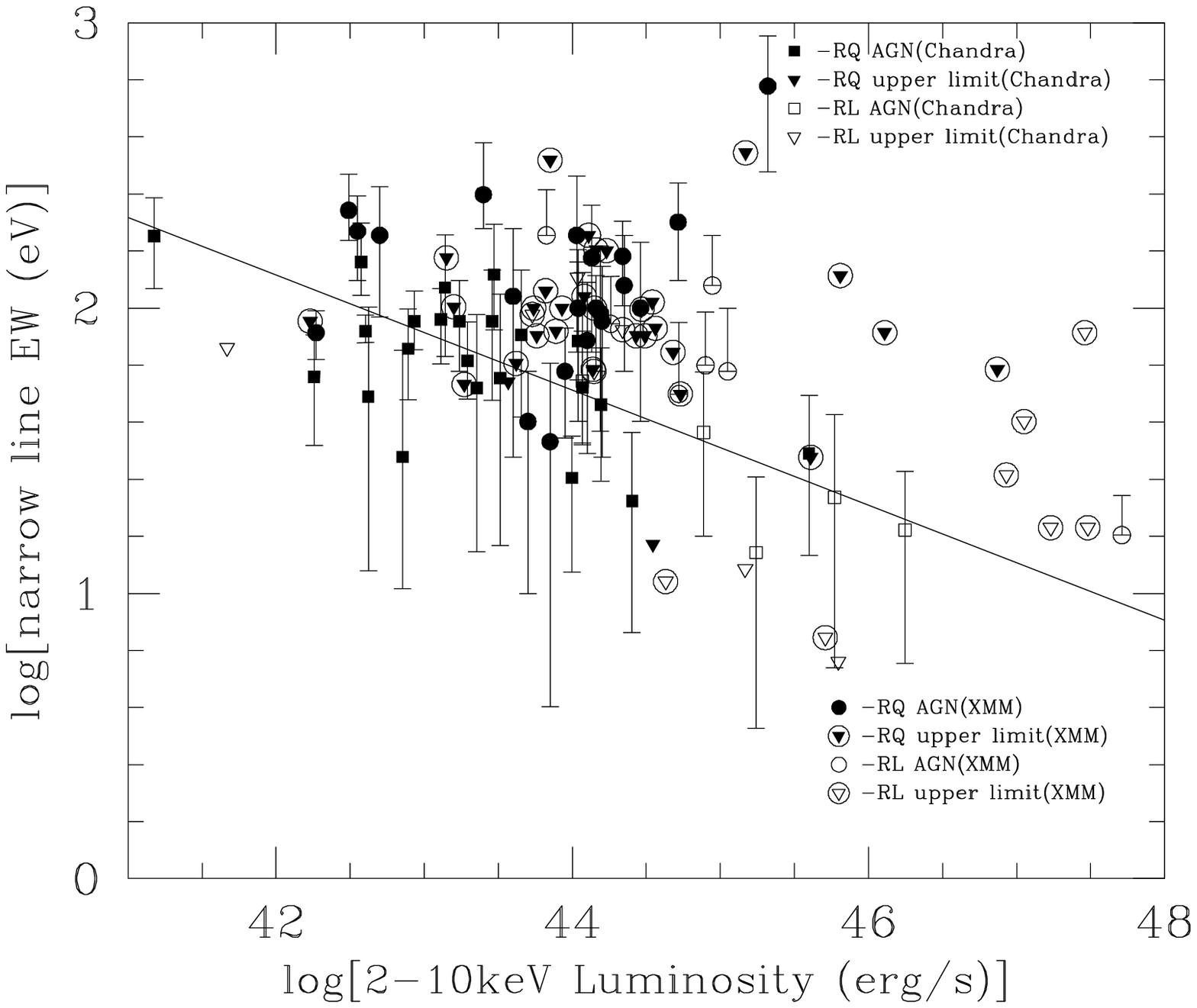}{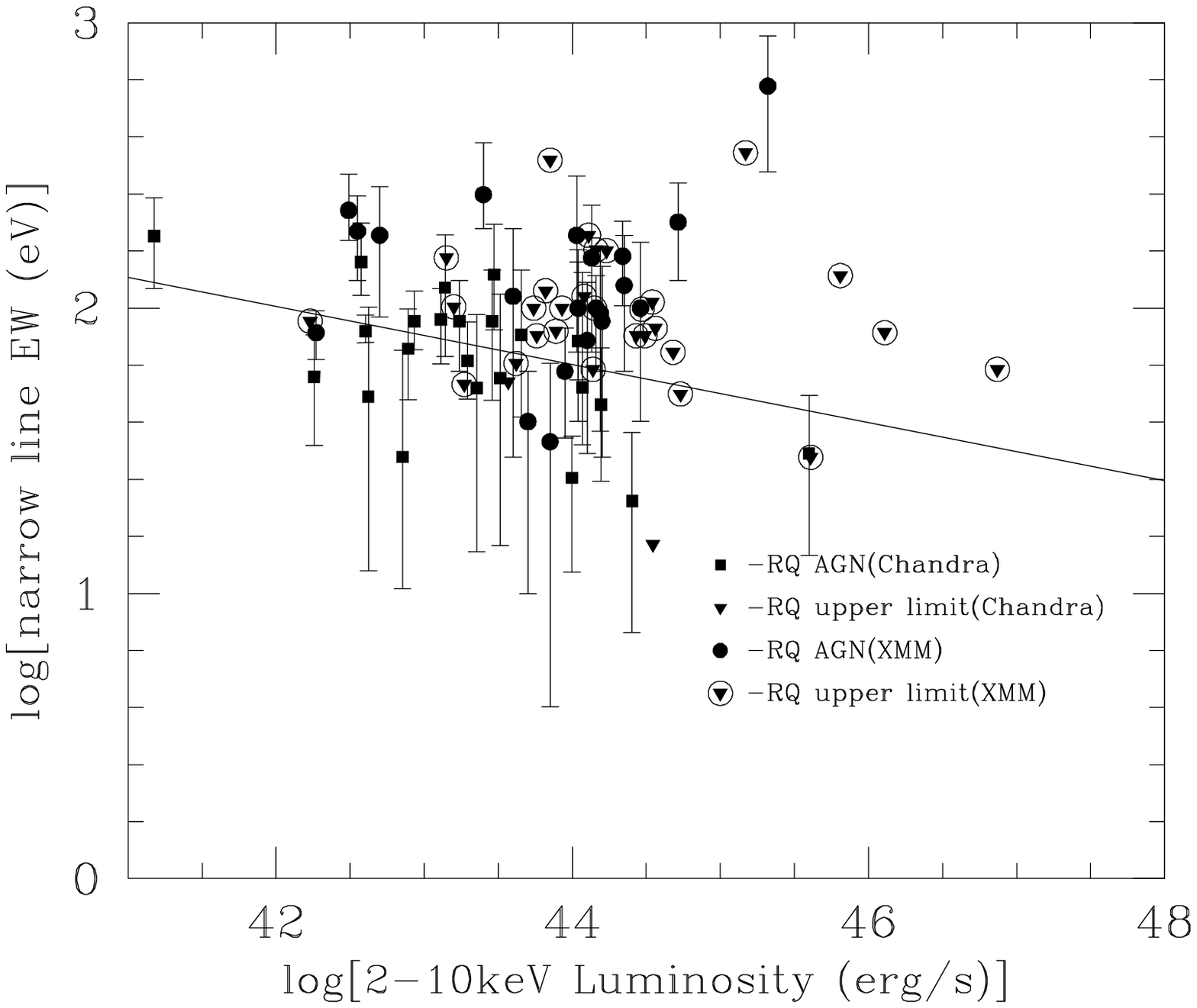}
\caption{The correlation between the luminosity and the EW for the 
combined sample. 
Left: including both radio quiet sources and radio loud sources;
Right: including radio quiet sources only. The solid lines show the
best-fit anti-correlation slopes. 
\label{fig4}}
\end{figure}

\clearpage

\begin{figure}
\plottwo{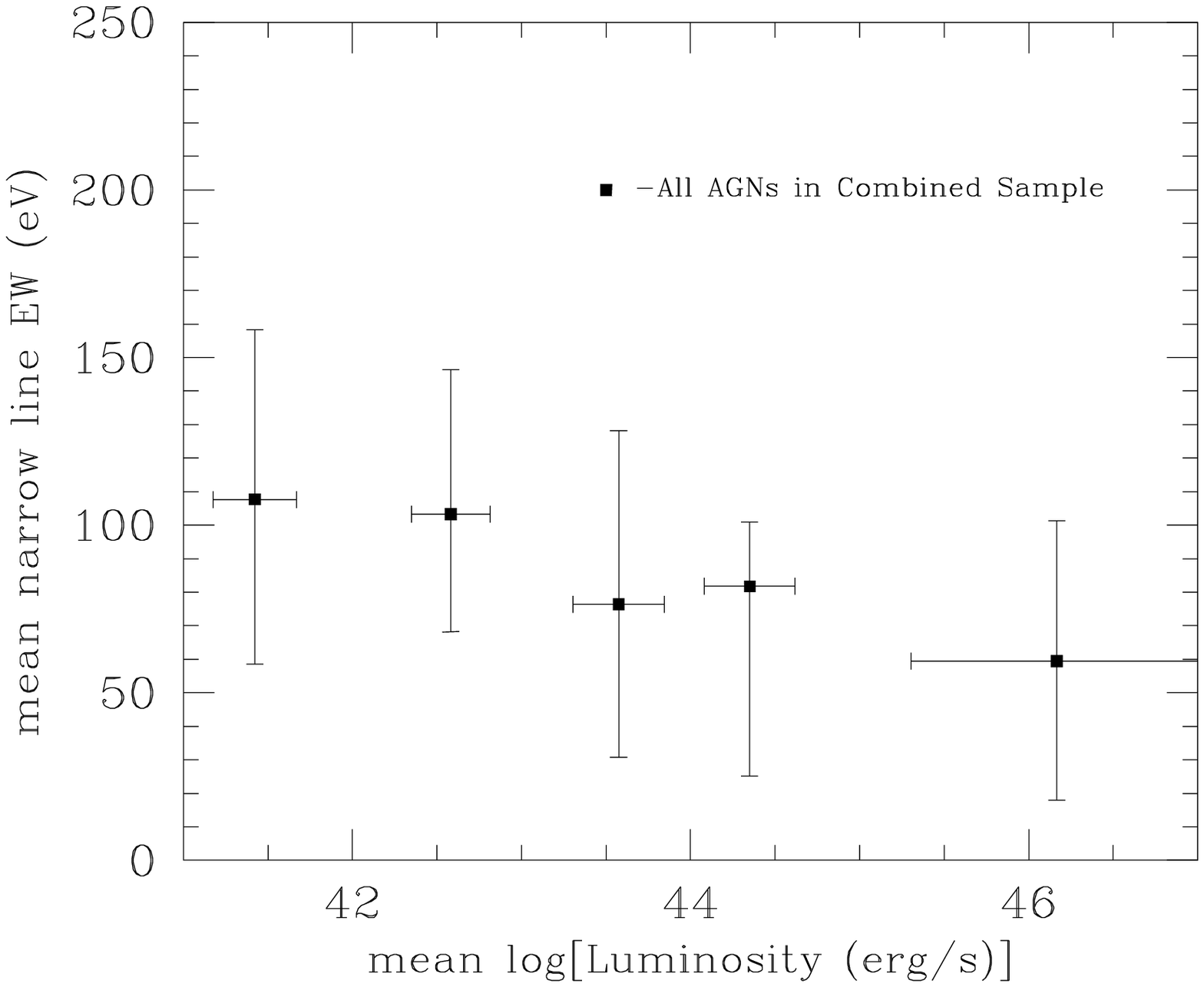}{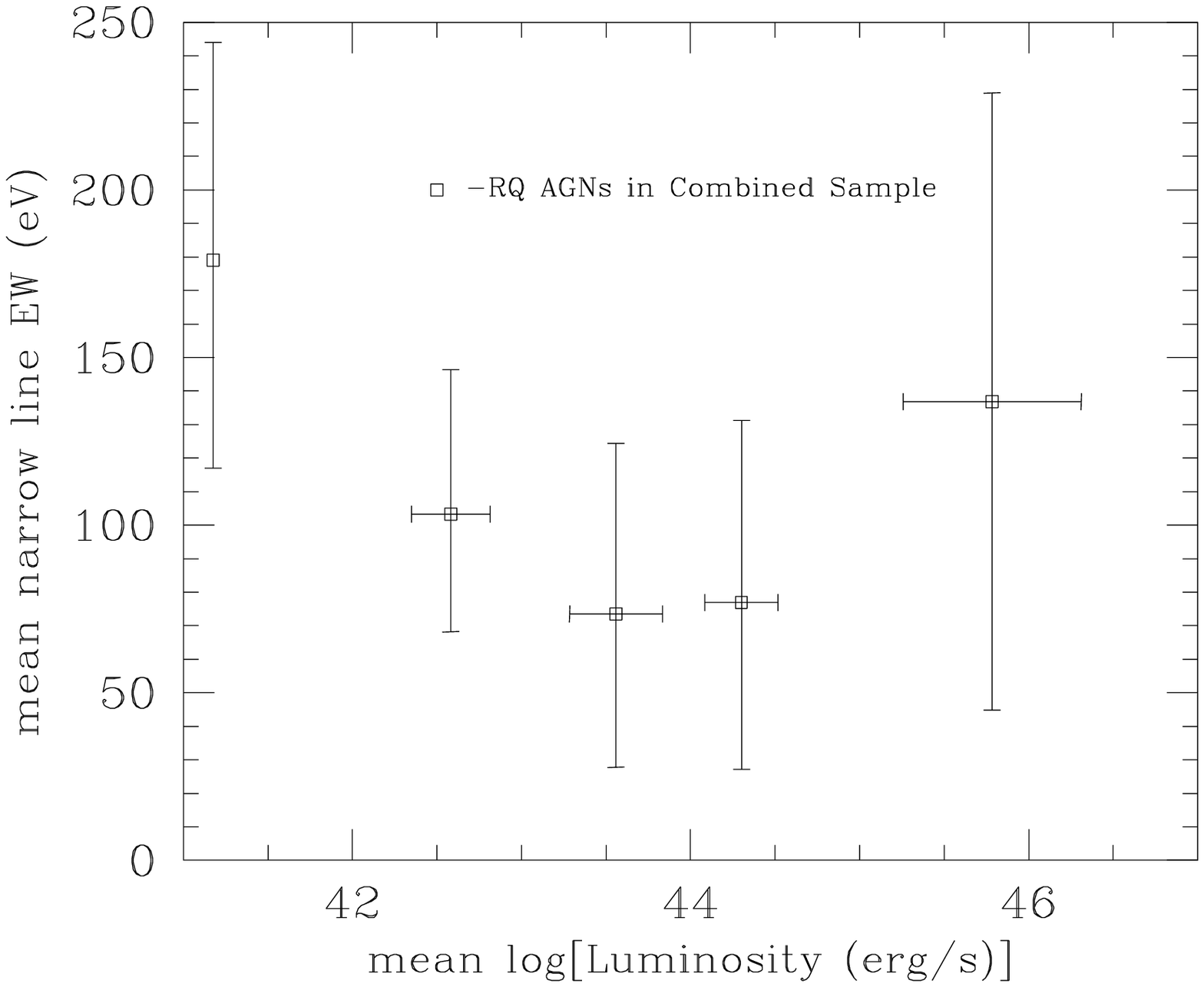}
\caption{Averaging the narrow line equivalent widths within 5 
luminosity bins for combined sample.
Left: All AGNs in combined sample;
Right: Only RQ AGNs in combined sample.
\label{fig5}}
\end{figure}

\clearpage

\begin{figure}
\plottwo{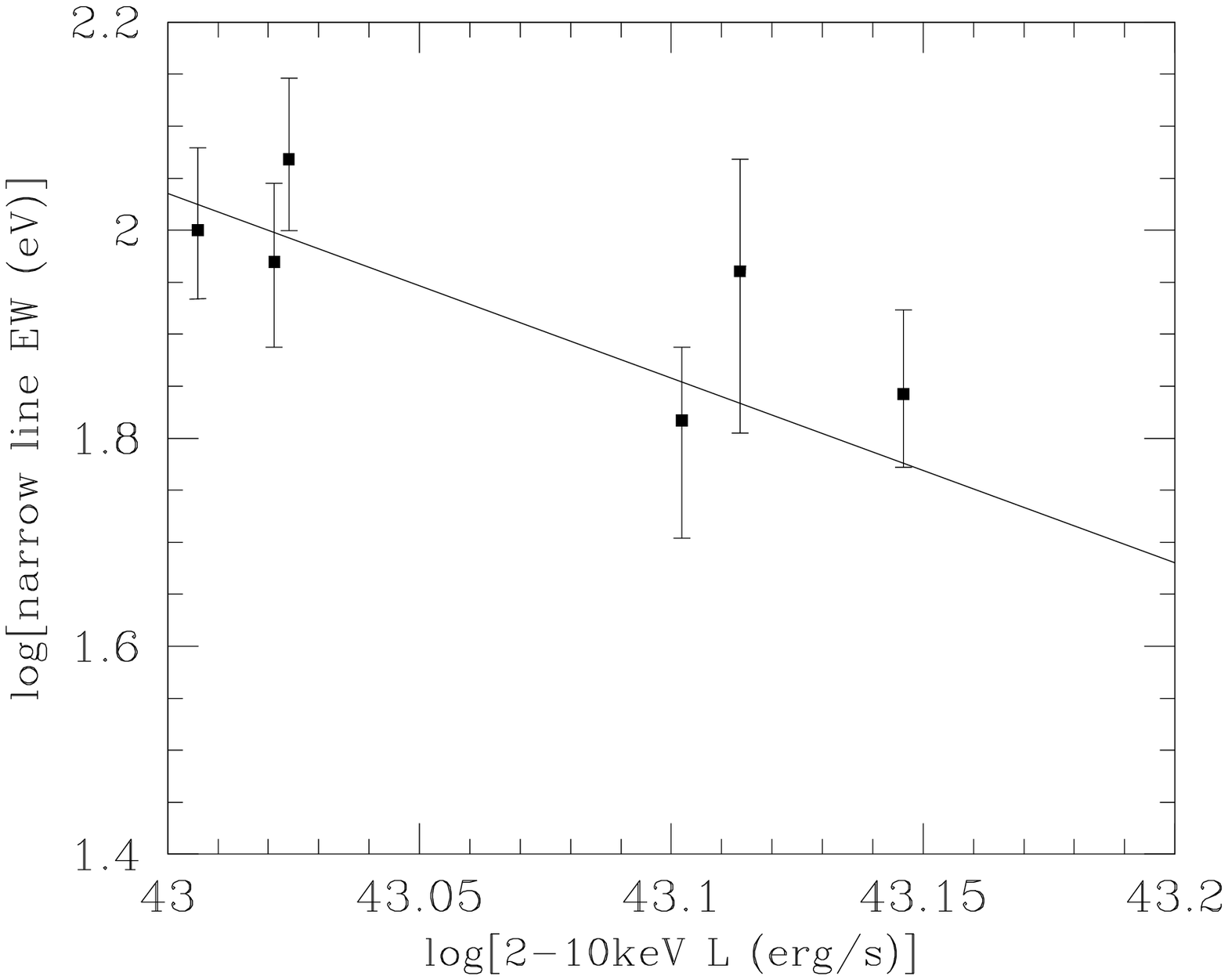}{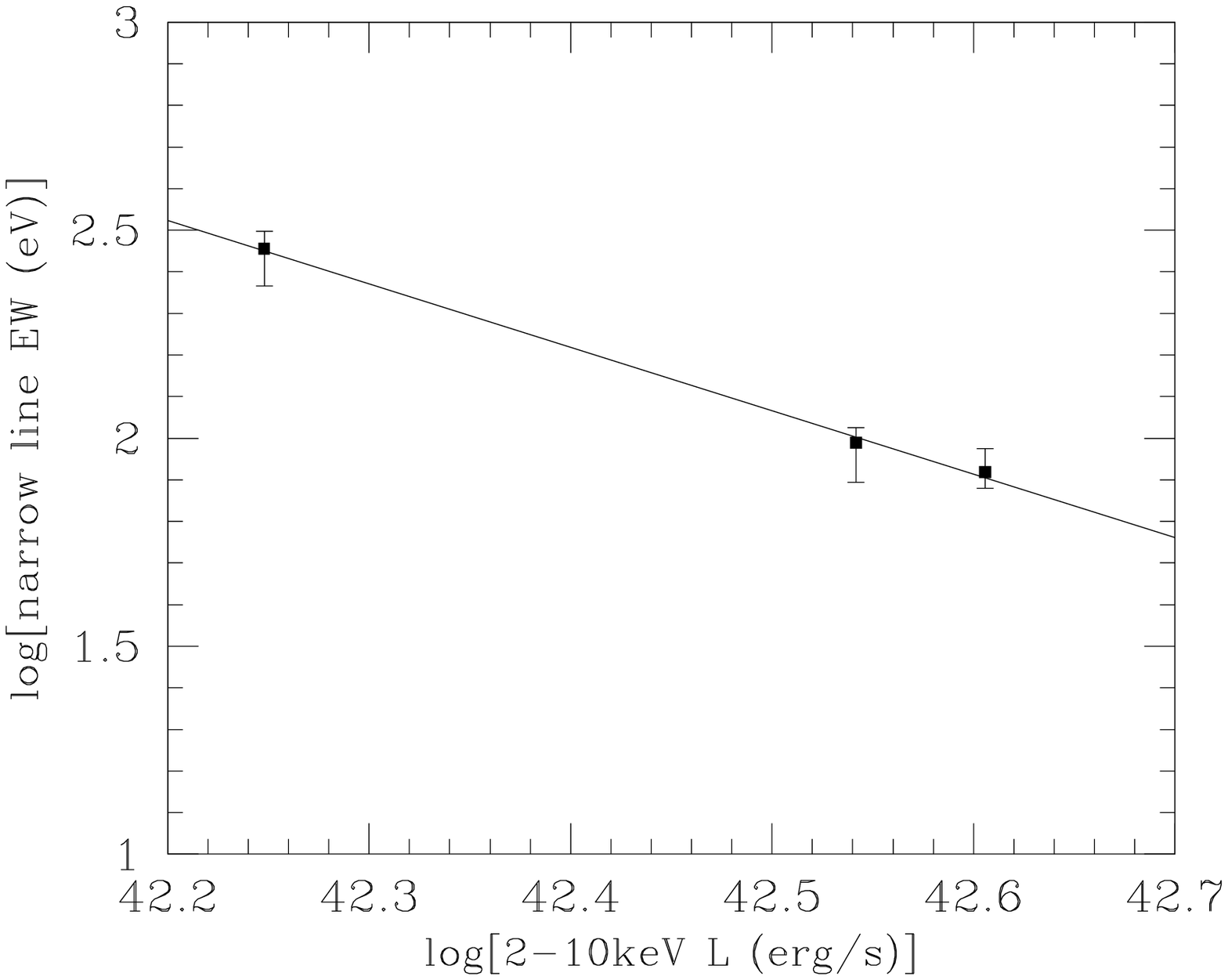}
\caption{Left: Six {\it Chandra} HETG observations of source NGC 3783. 
Right: Three {\it Chandra} HETG observations of source NGC 4151. 
The solid lines show the best-fit anti-correlation slopes.
See text for details.
\label{fig6}}
\end{figure}

\clearpage

\begin{figure}
\epsscale{.80}
\plotone{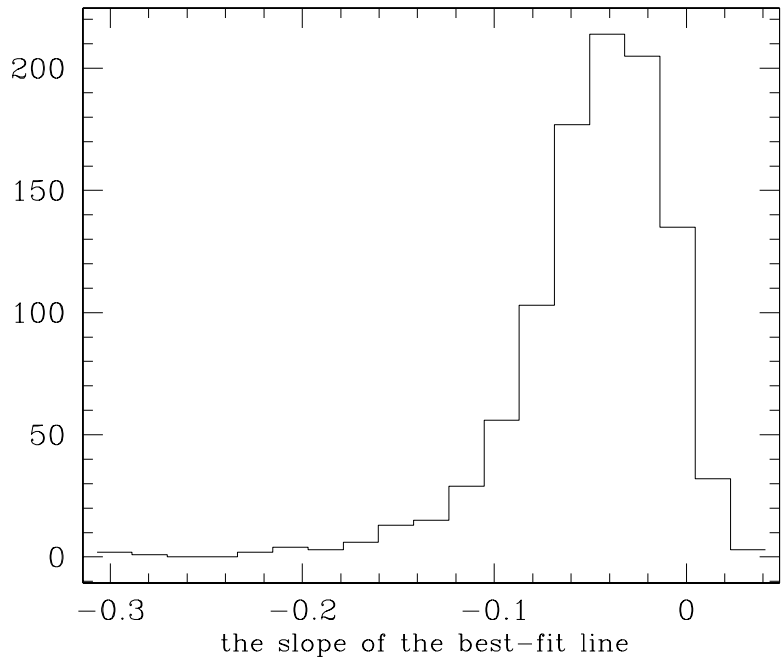}
\caption{The distribution of the best-fit power-law slopes of the 
artificial samples. 
\label{fig7}}
\end{figure}

\clearpage

\begin{deluxetable}{lcccccccccc}
\tabletypesize{\tiny}
\rotate
\tablecaption{Parameters of the Narrow Fe K$\alpha$ Emission Line from
 {\it Chandra} (HEG) Data \label{tbl-1}}
\tablewidth{0pt}
\tablehead{
\colhead{Source} & \colhead{$Z^a$} & \colhead{Spectral Type} &
\colhead{Pho.Index} & \colhead{$N_{H(Gal)}^b$} &
\colhead{$N_H^c$} & \colhead{$E^d$} &
\colhead{$FWHM^e$} & \colhead{$EW^f$} &
\colhead{$L^g$} & \colhead{C/bins} \\
\colhead{ } & \colhead{ } & \colhead{ } & \colhead{ } &
\colhead{($10^{22}$ $cm^{-2}$)} & \colhead{($10^{22}$ $cm^{-2}$)}
 & \colhead{(keV)} & \colhead{(km $s^{-1}$)} & \colhead{(eV)}
 & \colhead{(log[ ergs $s^{-1}$])} & \colhead{}
}
\startdata
NGC 4051 & 0.002 & RQ & 1.513 & 0.013 & h & 6.420(6.391-6.447) &
6160(3180-10870) & 179(117-244) & 41.176 & 284/283 \\
NGC 4151 & 0.003 & RQ & 1.406 & 0.020 & 3.265 & 6.389(6.385-6.394) &
2510(1970-3160) & 83.1(75.9-94.7) & 42.605 & 350/283 \\
NGC 7314 & 0.005 & RQ & 1.517 & 0.014 &   & 6.412(6.399-6.423) &
1000f & 57.7(33.2-80.9) & 42.256 & 317/283 \\
NGC 5506 & 0.007 & RQ & 1.751 & 0.037 & 3.282 & 6.411(6.395-6.418) &
3350(1240-4480) & 90.5(72.0-116) & 42.933 & 245/283 \\
$MCG-6-30-15^i$ & 0.008 & RQ & ... & ... & ... & 6.408(6.349-6.454) &
3250(0-12670) & 49(12-101) & 42.623 & ... \\
NGC 4593 & 0.009 & RQ & 1.689 & 0.023 &   & 6.403(6.390-6.410) &
2200(0-5670) & 72.6(48.1-100) & 42.892 & 290/283 \\
NGC 3783 & 0.009 & RQ & 1.668 & 0.083 & 0.874 & 6.398(6.387-6.410) &
2420(840-4010) & 92.1(64.4-118) & 43.114 & 318/283 \\
NGC 3516 & 0.009 & RQ & 0.866 & 0.029 & 1.171 & 6.398(6.383-6.417) &
4320(3030-7020) & 146(112-201) & 42.574 & 301/283 \\
MKN 766 & 0.012 & RQ & 1.950 & 0.017 &   & 6.421(6.393-6.457) &
1000f & 30.5(10.5-71.9) & 42.855 & 264/283 \\
IC 4329a & 0.016 & RQ & 1.681 & 0.044 & 0.232 & 6.404(6.388-6.415) &
1000f & 25.8(12.1-36.1) & 43.996 & 297/283 \\
NGC 5548 & 0.017 & RQ & 1.589 & 0.017 &   & 6.399(6.386-6.414) &
1690(0-2980) & 66.5(48.7-90.9) & 43.292 & 283/283 \\
NGC 7469 & 0.017 & RQ & 1.804 & 0.049 &   & 6.391(6.376-6.401) &
1100(0-4200) & 91.3(61.0-127) & 43.238 & 289/283 \\
NGC 526a & 0.019 & RQ & 1.081 & 0.022 & 0.046 & 6.405(6.383-6.423) &
1000f & 53.5(14.3-96.8) & 43.355 & 317/283 \\
ARK 564 & 0.025 & RQ & 2.388 & 0.063 &   & 6.4f &
1000f & $\lesssim$ 56.5 & 43.568 & 252/283 \\
MKN 590 & 0.027 & RQ & 1.704 & 0.027 & 0.400 & 6.4f &
1000f & 121(69.6-186) & 43.140 & 269/283 \\
MKN 290 & 0.030 & RQ & 1.700 & 0.017 & 0.845 & 6.429(6.378-6.521) &
3900(0-43540) & 58.5(15.1-115) & 43.513 & 242/283 \\
MKN 279 & 0.031 & RQ & 1.616 & 0.018 &   & 6.423(6.393-6.458) &
5240(3270-8130) & 135(87-203) & 43.472 & 322/283 \\
MKN 509 & 0.035 & RQ & 1.621 & 0.041 & 0.016 & 6.426(6.403-6.448) &
2860(0-6550) & 47.4(25.6-75.2) & 44.195 & 236/283 \\
NGC 985 & 0.043 & RQ & 1.581 & 0.029 & 1.317 & 6.387(6.3728-6.410) &
1000f & 84.0(43.4-142) & 43.656 & 298/283 \\
ESO 198-G24 & 0.045 & RQ & 1.498 & 0.045 &   & 6.387(6.372-6.400) &
1000f & 94.1(49.6-142) & 43.459 & 346/283 \\
F9 & 0.046 & RQ & 1.688 & 0.033 &   & 6.409(6.385-6.432) &
3780(2310-7240) & 80.1(58.7-152) & 44.039 & 256/283 \\
MRC 2251-178 & 0.064 & RQ & 1.517 & 0.027 & 0.407 & 6.4f &
1000f & 22.5(7.76-38.9) & 44.404 & 315/283 \\
IRAS 13349+2438 & 0.107 & RQ & 2.298 & 0.012 & 2.306 & 6.417(6.365-6.438) &
2940(230-8270) & 58.5(36.6-147) & 44.070 & 313/283 \\
PDS 456 & 0.184 & RQ & 1.854 & 0.200 & 1.049 & 6.4f & 1000f & $\lesssim$ 17.6 & 44.544 & 373/283 \\
H 1821+643 & 0.297 & RQ & 1.904 & 0.041 & 3.107 & 6.4f &
1000f & 40.1(17.6-64.1) & 45.598 & 290/283 \\
NGC 3227 & 0.003 & RL & 1.389 & 0.021 & 1.833 & 6.4f & 1000f &
$\lesssim$ 72.9 & 41.670 & 318/283 \\
MKN 421 & 0.031 & RL & 2.992 & 0.014 &   & 6.4f & 1000f &
$\lesssim$ 132 & 44.034 & 350/283 \\
3C 120 & 0.033 & RL & 1.636 & 0.111 & 0.055 & 6.418(6.404-6.428) &
1410(0-3810) & 57.2(34.8-85.8) & 44.067 & 305/283 \\
4C 74.26 & 0.104 & RL & 1.958 & 0.121 & 2.977 & 6.259(6.243-6.271) &
1000f & 40.4(17.6-65.8) & 44.886 & 301/283 \\
PKS 2155-304 & 0.116 & RL & 2.937 & 0.017 & 1.160 & 6.4f &
1000f & $\lesssim$ 13.6 & 45.168 & 277/283 \\
H 1426+428 & 0.129 & RL & 2.226 & 0.032 &   & 6.4f & 1000f & 15.7(3.79-28.9) &
45.241 & 265/283 \\
3C 273 & 0.158 & RL & 1.796 & 0.018 & 0.261 & 6.4f & 1000f & $\lesssim$ 6.69 & 45.795 & 297/283 \\
1 ES 1028+511 & 0.361 & RL & 2.574 & 0.012 & 2.340 & 6.366(6.310-6.381) &
1000f & 29.5(7.46-57.6) & 45.770 & 275/283 \\
3C 279 & 0.538 & RL & 1.740 & 0.022 & 2.097 & 6.564(6.548-6.584) & 1000f &
25.7(8.72-41.2) & 46.248 & 285/283 \\
\enddata
\tablecomments{Table \ref{tbl-1} All parameters are quoted in the source 
rest frame. Statistical errors are for the 90\% confidence level.}
\tablenotetext{a}{Redshifts obtained from {\it Quasars and Active  Galactic
 Nuclei (11th Ed.)} (Veron+. 2003)}
\tablenotetext{b}{Galactic column density}
\tablenotetext{c}{intrinsic absorption}
\tablenotetext{d}{Gaussian line center energy}
\tablenotetext{e}{FWHM, rounded to 10 km $s^{-1}$}
\tablenotetext{f}{Emission line equivalent width}
\tablenotetext{g}{2.--10. keV source frame luminosity}
\tablenotetext{h}{the intrinsic absorption value $\lesssim 10^{-5}$ and 
we ignored intrinsic absorption in our model}
\tablenotetext{i}{this fitting result was quoted from Yaqoob's Paper}
\end{deluxetable}

\clearpage
\begin{deluxetable}{lccccc}
\tabletypesize{\scriptsize}
\tablecaption{Parameters of the Narrow Fe K$\alpha$ Emission Line from
 {\it XMM-Newton} \label{tbl-2}}
\tablewidth{0pt}
\tablehead{
\colhead{Source} & \colhead{Redshift} & \colhead{Spectral Type} &
\colhead{EW} & \colhead{Luminosity} & \colhead{Ref.} \\
\colhead{ } & \colhead{ } & \colhead{ } & \colhead{(eV)} &
\colhead{(log[ ergs $s^{-1}$])} & \colhead{ }
}
\startdata
NGC 4151 & 0.003 & RQ & 187(184-190) & 42.27 & 1 \\
NGC 5506 & 0.006 & RQ & 70(50-90) & 42.83 & 1 \\
MCG-6-30-15 & 0.008 & RQ & 52(42-62) & 42.90 & 1 \\
NGC 3516 & 0.009 & RQ & 196(174-218) & 42.39 & 1 \\
NGC 4593 & 0.009 & RQ & 98(77-119) & 43.07 & 1 \\
Mrk 766 & 0.013 & RQ & 45(10-80) & 43.16 & 1 \\
IC 4329a & 0.016 & RQ & 30(18-42) & 44.77 & 1 \\
Mrk 359 & 0.017 & RQ & 220(146-294) & 42.49 & 1 \\
Mrk 1044 & 0.017 & RQ & 186(125-247) & 42.55 & 1 \\
NGC 5548 & 0.017 & RQ & 59(53-65) & 43.39 & 1 \\
Mrk 335 & 0.026 & RQ & $\lesssim$ 54 & 43.27 & 1 \\
Mrk 896 & 0.026 & RQ & 180(93-267) & 42.70 & 1 \\
Mrk 493 & 0.031 & RQ & $\lesssim$ 101 & 43.20 & 1 \\
Mrk 509 & 0.034 & RQ & 85(28-142) & 44.68 & 1 \\
Mrk 841 & 0.036 & RQ & $\lesssim$ 83 & 43.89 & 1 \\
1H 0707-495 & 0.041 & RQ & $\lesssim$ 90 & 42.23 & 1 \\
ESO 198-G24 & 0.046 & RQ & 104(55-153) & 43.27 & 1 \\
Fairall 9 & 0.047 & RQ & 139(113-165) & 43.97 & 1 \\
Mrk 926 & 0.047 & RQ & $\lesssim$ 61 & 44.14 & 1 \\
Ton S180 & 0.062 & RQ & $\lesssim$ 64 & 43.62 & 1 \\
MR 2251-178 & 0.064 & RQ & $\lesssim$ 74 & 44.46 & 1 \\
Mrk 304 & 0.066 & RQ & $\lesssim$ 115 & 43.82 & 1 \\
Mrk 205 & 0.071 & RQ & 60(35-85) & 43.95 & 1 \\
HE 1029-1401 & 0.086 & RQ & $\lesssim$ 105 & 44.54 & 1 \\
Mrk 1383 & 0.086 & RQ & 77(31-123) & 44.10 & 1 \\
1H 0419-577 & 0.104 & RQ & $\lesssim$ 85 & 44.56 & 1 \\
Mrk 876 & 0.129 & RQ & 96(37-155) & 44.19 & 1 \\
Q 0056-363 & 0.162 & RQ & $\lesssim$ 159 & 44.23 & 1 \\
PDS 456 & 0.184 & RQ & $\lesssim$ 18 & 44.77 & 1 \\
Q 0144-3938 & 0.244 & RQ & 152(102-202) & 44.34 & 1 \\
UM 269 & 0.308 & RQ & $\lesssim$ 99 & 44.47 & 1 \\
PB 05062 & 1.77 & RQ & $\lesssim$ 61 & 46.87 & 1 \\
1 Zw 1 & 0.061 & RQ & 34(4-64) & 43.85 & 2 \\
Akn 564 & 0.025 & RQ & $\lesssim$ 41 & 43.50 & 2 \\
NGC 4051 & 0.002 & RQ & 93(82-104) & 41.39 & 2 \\
NGC 7314 & 0.005 & RQ & 42(22-62) & 42.28 & 2 \\
Mrk 279 & 0.031 & RQ & 71(46-96) & 43.50 & 2 \\
NGC 3783 & 0.010 & RQ & 120(106-134) & 43.03 & 2 \\
NGC 7213 & 0.006 & RQ & 82(66-98) & 42.27 & 2 \\
NGC 7469 & 0.016 & RQ & 105(80-130) & 43.17 & 2 \\
SBS 0909+532 & 1.376 & RQ & 200(125-275) & 44.71 & 4 \\
PG 0157+001 & 0.163 & RQ & $\lesssim$ 330 & 43.85 & 3 \\
PG 0804+761 & 0.100 & RQ & 100(40-170) & 44.46 & 3 \\
PG 0844+349 & 0.064 & RQ & $\lesssim$ 100 & 43.74 & 3 \\
PG 0947+396 & 0.206 & RQ & 120(60-180) & 44.35 & 3 \\
PG 0953+414 & 0.234 & RQ & $\lesssim$ 50 & 44.73 & 3 \\
PG 1048+342 & 0.167 & RQ & 100(40-160) & 44.04 & 3 \\
PG 1114+445 & 0.144 & RQ & 100(60-130) & 44.16 & 3 \\
PG 1115+080 & 1.722 & RQ & $\lesssim$ 130 & 45.81 & 3 \\
PG 1115+407 & 0.154 & RQ & $\lesssim$ 100 & 43.93 & 3 \\
PG 1116+215 & 0.177 & RQ & $\lesssim$ 80 & 44.49 & 3 \\
PG 1202+281 & 0.165 & RQ & $\lesssim$ 80 & 44.43 & 3 \\
PG 1206+459 & 1.158 & RQ & $\lesssim$ 350 & 45.17 & 3 \\
PG 1211+143 & 0.081 & RQ & 40(10-60) & 43.70 & 3 \\
PG 1216+069 & 0.331 & RQ & $\lesssim$ 70 & 44.68 & 3 \\
PG 1244+026 & 0.048 & RQ & $\lesssim$ 150 & 43.15 & 3 \\
PG 1307+085 & 0.155 & RQ & $\lesssim$ 110 & 44.08 & 3 \\
PG 1322+659 & 0.168 & RQ & 180(70-290) & 44.03 & 3 \\
PG 1352+183 & 0.152 & RQ & 150(70-230) & 44.13 & 3 \\
PG 1402+261 & 0.164 & RQ & $\lesssim$ 100 & 44.15 & 3 \\
PG 1407+265 & 0.940 & RQ & $\lesssim$ 30 & 45.61 & 3 \\
PG 1411+442 & 0.090 & RQ & 250(190-380) & 43.40 & 3 \\
PG 1415+451 & 0.114 & RQ & 110(30-190) & 43.60 & 3 \\
PG 1427+480 & 0.221 & RQ & 90(30-140) & 44.20 & 3 \\
PG 1440+356 & 0.079 & RQ & $\lesssim$ 80 & 43.76 & 3 \\
PG 1444+407 & 0.267 & RQ & $\lesssim$ 180 & 44.11 & 3 \\
PG 1626+554 & 0.133 & RQ & $\lesssim$ 160 & 44.16 & 3 \\
PG 1630+377 & 1.476 & RQ & 600(300-900) & 45.32 & 3 \\
PG 1634+706 & 1.334 & RQ & $\lesssim$ 82 & 46.11 & 3 \\
NGC 3227 & 0.004 & RL & 191(168-214) & 41.37 & 1 \\
3c 120 & 0.033 & RL & 54(31-77) & 43.97 & 2 \\
PKS 1637-77 & 0.043 & RL & $\lesssim$ 95 & 43.73 & 1 \\
3 Zw 2 & 0.089 & RL & $\lesssim$ 84 & 44.34 & 1 \\
PKS 0558-504 & 0.137 & RL & $\lesssim$ 11 & 44.63 & 1 \\
3C 273 & 0.158 & RL & $\lesssim$ 9 & 45.72 & 1 \\
B2 1028+31 & 0.178 & RL & 88(43-129) & 44.26 & 1 \\
B2 1721+34 & 0.206 & RL & 63(29-97) & 44.90 & 1 \\
B2 1128+31 & 0.289 & RL & 50(11-89) & 44.72 & 1 \\
S5 0836+71 & 2.172 & RL & 16(10-22) & 47.71 & 1 \\
PKS 2149-30 & 2.345 & RL & $\lesssim$ 40 & 47.05 & 1 \\
PKS 0438-43 & 2.852 & RL & $\lesssim$ 26 & 46.93 & 1 \\
PKS 0537-286 & 3.104 & RL & $\lesssim$ 82 & 47.46 & 1 \\
PKS 2126-15 & 3.268 & RL & $\lesssim$ 17 & 47.48 & 1 \\
S5 0014+81 & 3.366 & RL & $\lesssim$ 17 & 47.23 & 1 \\
PG 0007+106 & 0.089 & RL & $\lesssim$ 60 & 44.146 & 3 \\
PG 1100+772 & 0.312 & RL & 60(20-100) & 45.049 & 3 \\
PG 1226+023 & 0.158 & RL & $\lesssim$ 7 & 45.708 & 3 \\
PG 1309+355 & 0.184 & RL & 180(140-260) & 43.826 & 3 \\
PG 1512+370 & 0.371 & RL & 120(60-180) & 44.946 & 3 \\
\enddata
\tablecomments{The luminosities were calculated 
for the 2.--10. keV rest frame.} 
\tablerefs{(1) Page et al. (2004a); 
(2) Zhou \& Wang (2005); 
(3) Jim$\acute{\rm e}$nez-Bail$\acute{\rm o}$n et al. (2005);
(4) Page et al. (2004b)} 
\end{deluxetable}

\clearpage

\begin{deluxetable}{lccc}
\tabletypesize{\scriptsize}
\tablecaption{Results of Linear Regression \label{tbl-3}}
\tablewidth{0pt}
\tablehead{
\colhead{Sample} & \colhead{Number of objects} & \colhead{Slope coefficient} &
\colhead{Spearman's coefficient}
}
\startdata
Chandra (RQ+RL) & 34 & $-0.1940\pm0.0332$ & -0.651 \\
Chandra (RQ) & 25 & $-0.1759\pm0.0493$ & -0.520 \\
XMM (RQ+RL) & 89 & $-0.2097\pm0.0503$ & -0.463 \\
XMM (RQ) & 69 & $-0.1023\pm0.0558$ & -0.280 \\
Combined (RQ+RL) & 101 & $-0.2015\pm0.0426$ & -0.469 \\
Combined (RQ) & 75 & $-0.1019\pm0.0524$ & -0.266 \\
\enddata
\end{deluxetable}
\end{document}